\documentclass[letterpaper]{JHEP3}
\usepackage{epsfig}
\usepackage{amsmath}

\newcommand{\tht}{\tilde\theta}

\def\to{\rightarrow}
\def\be{\begin{equation}}
\def\ee{\end{equation}}
\def\bea{\begin{eqnarray}}
\def\eea{\end{eqnarray}}

\def\half{\frac12}
\def\cn{${\bf C}/{\bf Z}_N$ }

\def\a{\alpha}

\def\d{\delta}

\def\f{\phi}

\def\k{\kappa}

\def\m{\mu}

\def\p{\pi}

\def\s{\sigma}

\title{The large $N$ limit of \cn and supergravity}

\author{Matthew Headrick \\
Center for Theoretical Physics, Massachusetts Institute of Technology \\
77 Massachusetts Ave., Cambridge MA 02139, USA \\
E-mail: \email{headrick@mit.edu}}

\author{Joris Raeymaekers \\
School of Physics, Korea Institute for Advanced Study \\
207-43 Cheongnyangni 2-dong, Dongdaemun-Gu, Seoul 130-722, Korea\\
E-mail: \email{joris@kias.re.kr}}

\abstract{
The \cn orbifold of type II string theory has localized tachyons with $m^2$ ranging from $-1+1/N$ to $-2/N$ in units of $2/\alpha'$. We show that by restricting attention to the lightest tachyons it is possible to take a zero-slope limit where $N$ is taken to infinity while $N\alpha'$ is held fixed. This is done by applying Buscher duality in the angular direction of the cone to obtain a supergravity solution on which the tachyons are gravitational instabilities. In this picture, supergravity provides a natural off-shell description of the tachyonic interactions. For example, the three-point couplings can be read off easily (to leading order in $1/N$) from the supergravity action, and are in agreement with the on-shell couplings computed using CFT techniques.}

\keywords{Tachyon Condensation, String Duality}

\preprint{\hepth{0411148} \\ KIAS-P04043 \\ MIT-CTP-3555}

\begin{document}

\section{Introduction}

Despite progress in recent years, understanding the physics of unstable closed-string backgrounds remains an important open problem. There are two classes of instabilities that one encounters which are qualitatively quite distinct. The first class contains the usual gravitational instabilities such as the Gregory-Laflamme instability of black strings; such instabilities are common to all theories of gravity and exist already in the supergravity approximation to string theory. The second class of instabilities are more stringy in nature and have to do with the fact that most non-supersymmetric string backgrounds contain tachyons in their perturbative spectra, typically with string scale masses.

The difficulties in analyzing the dynamics of these instabilities are both stringy and non-stringy in nature. The non-stringy difficulties are encountered already in the study of conventional gravitational instabilities and have to do with the fact that understanding these instabilities does not reduce to finding the minimum of some local potential energy density as is the case in non-gravitational field theories. The stringy difficulties, on the other hand, arise because stringy tachyons typically have masses of the order of the string scale. As a result, their condensation cannot be studied adequately within the low energy effective theory and necessarily requires a full-fledged off-shell formalism.

In point-particle theories, various instabilities have been analyzed quantitatively using off-shell effective field theory, and their physics concerning phase transitions and symmetry breaking is very well understood. Similarly, in open string theories, the tachyonic instabilities can be analyzed adequately within open string field theory and their physics has to do with the decay of various unstable D-brane configurations in the theory. For closed strings, on the other hand, the required off-shell formalism is considerably more complicated. Even for the bosonic string, the closed string field theory does not have the simple cubic form as in Witten's open string field theory, although a well-developed formalism with non-polynomial interactions is available \cite{Zwiebach:1993ie}. Recently a proposal has been made for a non-polynomial interacting heterotic string field theory \cite{Okawa:2004ii,Berkovits:2004xh}. For type II superstrings however, there is a string field theory formalism available only for the free theory \cite{Berkovits:1996tn}, but not yet for the interacting theory. As a result, progress in closed string tachyon condensation to date has for the most part been made using indirect arguments such as world-sheet renormalization group flow and D-brane probe analysis (see the reviews \cite{Martinec:2002tz,Headrick:2004hz} and references therein).

The purpose of this paper is to exhibit and exploit a simple stringy effect, namely T-duality, to tame the stringiness in one well-studied system containing closed-string tachyons, the \cn orbifold of type II string theory. We will show that certain tachyons of \cn at large $N$ are manifested in the dual background as gravitational instabilities. Furthermore, in the dual description supergravity\footnote{Strictly speaking the dual background is governed not by a supergravity but, as we will discuss, by the low energy effective action of type 0 string theory. However, since we will be concerned mostly with the massless NS-NS fields, whose action is the same as in supergravity, we will sometimes be sloppy and just use the word ``supergravity"
to mean that action.} provides a natural off-shell action from which one can read off the interactions of these tachyons. In particular, the duality will allow us to easily test an interesting conjecture about those interactions which was made in the paper \cite{Dabholkar:2004}.

This paper is organized as follows. In the rest of this introduction, we quickly review the relevant aspects of the \cn system, review the paper \cite{Dabholkar:2004}, and summarize the T-duality and dimensional argument which provide a shortcut for testing the conjecture made in that paper. In Section 2 we carefully derive the duality of both the background and the tachyons of interest. In Section 3 we illustrate how supergravity in the dual picture provides a natural off-shell action for the tachyons of interest by using it to calculate their self-interactions and interactions with massless fields, and show that the resulting S-matrix elements agree with those obtained by conformal field theory methods.

\subsection{Review of \cn}

The orbifold ${\bf R}^{1,7}\times{\bf C}/{\bf Z}_N$ is a solution of the classical equations of motion of type II string theory for any odd integer $N$ \cite{Dabholkar:1995ai,Lowe:1995ah}.
The \cn directions have the geometry of a cone with deficit angle $2\pi(1- 1/N)$, and the string spectrum on this background contains tachyons in the twisted sectors that are localized near the tip of the cone, the orbifold fixed point. The dynamics of the condensation of these tachyons
is not yet well understood quantitatively, but the following qualitative picture, originally advocated in \cite{Adams:2001sv}, has by now received substantial support:\footnote{The support comes both from indirect arguments such as D-brane probe analysis \cite{Adams:2001sv,Harvey:2001wm} and world-sheet RG flow \cite{Adams:2001sv,Harvey:2001wm,Vafa:2001ra,Gutperle:2002ki,Minwalla:2003hj} (see \cite{Headrick:2004hz} for a review), and from study of the analogous bosonic orbifold using closed string field theory \cite{Okawa:2004rh,Bergman:2004st}. Other papers on the \cn system include \cite{Dabholkar:2001if,Dabholkar:2001wn,Basu:2002jt,Nakamura:2003wz,Gregory:2003yb,Headrick:2003yu}.} The orbifold should be viewed as an unstable soliton inhabiting the same superselection sector as (the same theory in) flat space, and the condensation of its tachyons leads (generically) to flat space, or (with fine-tuning) to another \cn orbifold with a smaller value of $N$. In this sense the orbifold is similar to an unstable D-brane. An important difference is that a D-brane is a soliton of the open string fields and therefore has a tension that goes only like $g_{\rm s}^{-1}$, whereas the orbifold (like an NS5-brane) is a soliton of the closed string fields and therefore has a tension that goes like $g_{\rm s}^{-2}$:
\begin{equation}
T_{{\bf C}/{\bf Z}_N} = \frac{2\pi}{\kappa^2}\left(1-\frac1N\right),
\end{equation}
where $\kappa\propto g_{\rm s}$ is the gravitational coupling.

A few basic facts about the tachyons will be important for us. The orbifold has $N-1$ twisted sectors, labelled by their charges under the  quantum ${\bf Z}_N$ symmetry. Each twisted sector contains at least one tachyon; we will refer to the most tachyonic one with charge $k$ as $T_k$. Since the complex conjugate field $\bar T_k$ has charge $-k$, we can restrict $k$ to lie in the range $0<k<N/2$ without loss of generality. $T_k$ is a $7+1$ dimensional field living on the orbifold fixed plane, with mass
\begin{equation}\label{spectrum}
m_k^2 = -\frac2{N\alpha'}
\begin{cases}N-k, & \hbox{$k$ odd} \\ k, & \hbox{$k$ even}\end{cases}.
\end{equation}
While the result of dynamical condensation of this tachyon is, as noted above, conjectural, the endpoint of the world-sheet RG flow seeded by its vertex operator is known \cite{Vafa:2001ra}: it is the orbifold ${\bf C}/{\bf Z}_{k}$ (if $k$ is odd) or ${\bf C}/{\bf Z}_{N-k}$ (if $k$ is even).

\subsection{The DIR conjecture}

We see from (\ref{spectrum}) that at small $N$ the tachyons have string-scale masses---there is no other mass scale in the system. However, at large $N$ the lightest tachyons, namely those with $k$ even and much smaller than $N$, have masses parametrically smaller than the string scale. Dabholkar, Iqubal, and Raeymaekers (DIR) \cite{Dabholkar:2004} suggested that it might be meaningful to construct a low energy field theory for these modes, without resorting to the machinery of string field theory. Imagine for example that the potential energy for $T_k$ takes the following form:
\begin{equation}\label{potential}
V(T_k) = \frac{2\pi\alpha'}{\kappa^2}\left(m_k^2|T_k|^2 + \frac{\lambda_k}4|T_k|^4 \right)
\end{equation}
(the prefactor is for convenience and makes $T_k$ a dimensionless field; the same prefactor would multiply the kinetic term). For the moment we are neglecting both higher order terms in $|T_k|^2$ and terms that couple different tachyons to each other, such as $T_k^2\bar T_{2k}$. If $\lambda_k$ is positive, then this potential has a minimum at
\begin{equation}\label{potmin}
|T^{\rm min}_k|^2 = -\frac{2m_k^2}{\lambda_k}, \qquad
V^{\rm min} = -\frac{2\pi\alpha'm_k^4}{\kappa^2\lambda_k}.
\end{equation}
If, furthermore, $\lambda_k$ is finite in the limit $N\to\infty$, then at large $N$ the energy at this minimum is small, of order $N^{-2}$ in Planck units. In this situation it is consistent to treat the field with this low energy effective action, and to ignore higher order terms in $|T_k|^2$ (unless their coefficients are parametrically large in $N$). The situation is analogous to that of the Higgs field in the Standard Model, which, thanks to the mass hierarchy, can be satisfactorily analyzed using that low energy effective theory without knowing what the full theory at the Planck scale is.

Once the coupling to the metric is taken into account, the minimum (\ref{potmin}) would correspond to a static solution similar to \cn but with a slightly smaller deficit angle. This leads to the suggestion that it is a lower-order orbifold. In particular, if the world-sheet RG flow seeded by a particular relevant vertex operator indicates the result of condensation of the corresponding tachyon, then this solution should be identified with ${\bf C}/{\bf Z}_{N-k}$. The difference between the tensions of the two solitons is of the right order, $N^{-2}$:
\begin{equation}\label{tendiff}
T_{{\bf C}/{\bf Z}_N} - T_{{\bf C}/{\bf Z}_{N-k}} = \frac{2\pi}{\kappa^2}\left(-\frac{2k}{N^2} + \mathcal{O}\left(\frac1{N^3}\right)\right).
\end{equation}
Comparing (\ref{potmin}) to (\ref{tendiff}) yields a precise prediction for the value of $\lambda_k$ in the large $N$ limit:
\begin{equation}\label{lambdapredicted}
\lambda_k = \frac{2k}{\alpha'} + \mathcal{O}\left(\frac1N\right) \qquad {\rm (predicted)}.
\end{equation}
This scenario also predicts that the terms $|T_k|^6$ and higher are not enhanced by positive powers of $N$, and that mixed cubic terms of the form $T_k^2\bar T_{2k}$ are suppressed by at least one power of $N^{-1}$; otherwise they would invalidate the analysis that led to (\ref{potmin}).

This scenario was conjectured by the authors of \cite{Dabholkar:2004}, who proceeded to calculate $\lambda_k$ as well as the coefficient of $T_k^2\bar T_{2k}$ in string perturbation theory. They showed that the vertex $T_k^2\bar T_{2k}$ is forbidden (even at finite $N$) by world-sheet $H$-charge conservation, confirming the above prediction. Disappointingly, however, the result for $\lambda_k$ fell short of the prediction (\ref{lambdapredicted}): $\lambda_k\sim N^{-2}$.\footnote{Actually, an error led the authors of \cite{Dabholkar:2004} to report $\lambda_k\sim N^{-3}$. However, the difference is hardly significant given that non-linear field redefinitions can change $\lambda_k$ arbitrarily by terms of order $N^{-1}$ (the mass-squared of $T_k$). In other words, the order $N^{-1}$ part of $\lambda_k$ is not well defined until one fixes an off-shell extension of string perturbation theory.} The calculation (which required sophisticated techniques in orbifold conformal field theory and ran to 18 pages) was done by taking the four-point tachyon scattering amplitude and subtracting diagrams with an exchanged intermediate particle in order to extract the elementary quartic coupling. The very small final result depended on a cancellation between these two kinds of terms, each of order $N^{-1}$.

\subsection{A gravity dual}

Given that the tachyons of interest in this system are parametrically lighter than the string scale, it is natural to wonder if they can be studied using non-stringy methods that might be either technically or conceptually easier than the CFT methods employed in \cite{Dabholkar:2004}. More precisely, one can ask whether the \cn system admits a sensible zero-slope limit of the following form:
\begin{equation}\label{zeroslope}
\alpha'\to0, \qquad N\to\infty, \qquad N\alpha' \hbox{ fixed}.
\end{equation}
We will show that in fact it does, in the sense that all states whose masses remain finite in this limit have couplings that also remain finite (where the couplings are defined, as in (\ref{potential}), without the factor $2\pi\alpha'/\kappa^2$). The existence of this limit immediately establishes that $\lambda_k$ is (at most) of order $N^{-1}$, since dimensionally it must have a factor $\alpha^{\prime-1}$. This confirms the result of the CFT calculation of \cite{Dabholkar:2004}, showing that the DIR conjecture cannot be right.

Ironically, to show that \cn admits a sensible zero-slope limit of the form (\ref{zeroslope}) requires invoking a stringy effect, namely T-duality.\footnote{The potential utility of doing a T-duality in the angular direction in studying \cn at large $N$ was pointed out in \cite{Adams:2001sv}.} Here we will give a quick derivation, which we will repeat including details and caveats in the next section. At large $N$ the cone degenerates, and is best viewed as a circle fibered over a half-line, where the radius $R$ of the circle varies slowly as a function of the base point (figure 1, top):
\begin{equation}\label{fibersize}
R(r) = \frac rN,
\end{equation}
where $r$ is the proper distance to the fixed point. The tachyons are in twisted sectors, which means that they are wound around the circle. Their tension then keeps them close to the fixed point, where the circle is small; how close depends on how many times they are wound. As we will show in the next section, strings in twisted sector $k$ are typically found at a distance $\sqrt{N\alpha'/k}$ from the fixed point. For the tachyons of interest, which have $k\ll N$, this is many times the string length. But according to (\ref{fibersize}) the circle size at that distance is smaller than the string length, $R(\sqrt{N\alpha'/k})=\sqrt{\alpha'/Nk}$. Therefore if we're interested in what's going on in that region, we should T-dualize the fiber. The result is a geometry that is smooth in the limit (\ref{zeroslope}) and is characterized by the single length scale $\sqrt{N\alpha'}$ (figure 1, bottom). (The T-dual geometry has a boundary at $r=0$; we will be more careful about this in the next section.) The tachyons have winding but no momentum around the original circle, and are therefore momentum modes of the T-dual circle. They are so light that they cannot be excited string states, so they must simply be modes of the massless fields. It turns out that they are gravitons which, as in the Gregory--Laflamme instability of black strings, acquire a negative mass due to the curvature of the background. (The phenomenon of a localized closed string tachyon being T-dual in a certain limit to a gravitational instability was also seen in the context of Melvin models and twisted circles \cite{Russo:2001tf,David:2001vm}.)

The couplings among these tachyons can be computed to first order in $N^{-1}$ by expanding the supergravity action about the T-dual background. This is an exercise which is algebraically complicated but straightforward enough using {\it Mathematica}. In Section 3 below we will report the results and compare them to those obtained by conformal field theory methods. We will also
compute couplings between the tachyons and massless fields on ${\bf C}/{\bf Z}_N$, and even show that some couplings with the heaviest tachyons (those with $k$ odd and much smaller than $N$) can be computed by this method.

\EPSFIGURE{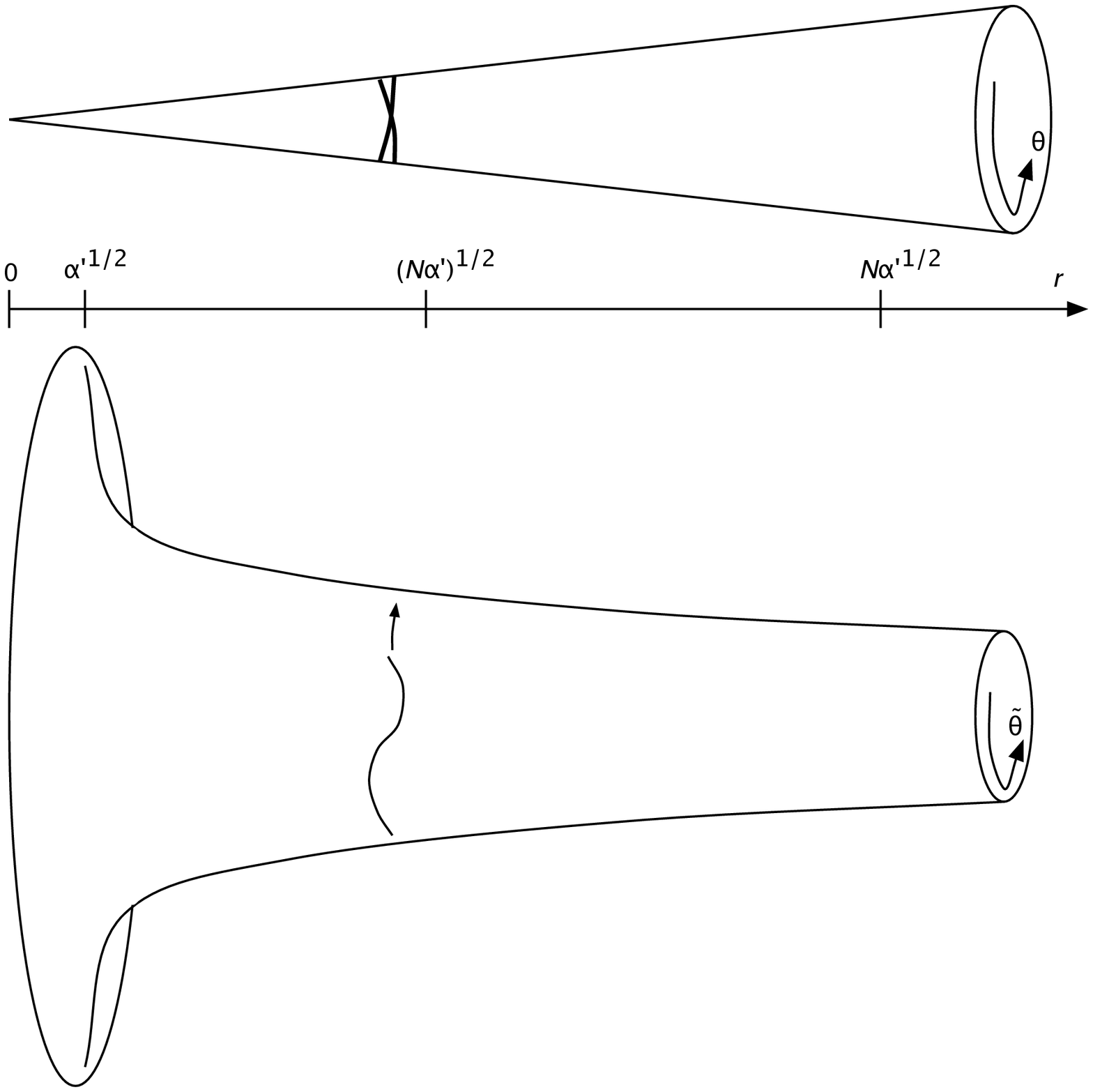,width=5.5in}{Schematic diagram of the relevant geometries and length scales, described briefly in subsection 1.3 and in detail in section 2. Top: at large $N$, it is useful to view the geometry of \cn as a circle (parametrized by $\theta$) fibered over a
half-line (parametrized by $r$), where the size of the circle varies slowly over the base.
Bottom: it is appropriate to describe the physics in the range of radii $\sqrt{\alpha'}<r<N\sqrt{\alpha'}$ in terms of the space obtained by T-dualizing the fiber, whose geometry (\ref{postt}) is characterized by the single length scale $\sqrt{N\alpha'}$. Strings with small twist number $k$ compared to $N$ have wave functions that are Gaussians in the radial direction of width $\sqrt{N\alpha'/k}$. In the T-dual picture, tachyons with even $k$ are gravitational instabilities, while tachyons with odd $k$ are modes of the type 0 bulk tachyon.}

\section{\cn tachyons as gravitational instabilities}

On ${\bf R}^{1,7}\times{\bf C}/{\bf Z}_N$ we use coordinates $x^\mu,r,\theta$ ($\mu=0,\dots,7$, $\theta\sim\theta+2\pi$), with metric
\begin{equation}\label{pret}
ds^2 = dx^2+dr^2 + \left(\frac rNd\theta\right)^2,
\end{equation}
The dilaton is a constant $\Phi_0$, and we will use $\kappa$ to denote the gravitational coupling corresponding to $g_{\rm s}=e^{\Phi_0}$.

For large $N$, the radius of the $\theta$ circle, $R = r/N$, is a slowly varying function of $r$, and in the region $r<N\sqrt{\alpha'}$ it is small in string units. Therefore, as discussed in \cite{Adams:2001sv}, it is useful to T-dualize the circle there, which we will do in the first subsection. In the next subsection we will show that the system's heaviest tachyons are modes of a bulk tachyon that appears in that background \cite{Adams:2001sv}, and that the lightest tachyons are gravitational instabilities of the T-dual geometry.

\subsection{The T-dual of \cn at large $N$}\label{thermalt}

Because the $\theta$ circle is a thermal circle (one on which the fermions have anti-periodic boundary conditions), we will begin by reviewing how T-duality works on thermal circles. The best way to define a thermal circle of radius $R$ in string theory is to consider it as a $Z_2$ orbifold of a normal circle of radius $2R$, by the operator $(-1)^{F+p}$. Here $F$ is the spacetime fermion number, while $p$ is the integral momentum along the circle, so that $(-1)^p$ rotates the circle through an angle $\pi$. The orbifold projection forces bosons to have even $p$ and fermions to have odd $p$. The twisted sector consists of strings that close only up to a rotation through an angle $\pi$, i.e.\ that have half-odd winding number; these have a reversed GSO projection, just like the twisted sector of the orbifold by $(-1)^F$ that produces type 0 strings. This is precisely the spectrum of a thermal circle of radius $R$, with the additional information that strings of
odd winding number have a reversed GSO projection.

To obtain the T-dual of this thermal circle, we apply the same orbifold after T-dualizing the normal circle of radius $2R$. The T-dual circle has radius $\alpha'/(2R)$, half as big as we might have
expected. In this picture $p$ represents the winding number. The orbifold projection, by $(-1)^{F+p}$, thus requires that strings of even winding number be bosons and of odd winding number be fermions. What are the twisted sector strings? Before the T-duality (on the circle of radius $2R$) they were strings with half-odd winding number and reversed GSO projection. After T-duality, they are therefore strings with half-odd momentum---i.e.\ anti-periodic boundary conditions---and reversed GSO projection.

If we are interested in a situation in which $R$ is small in string units, then in the T-dual picture we can neglect winding modes. The non-winding modes have the same field content as type 0 theory, but with anti-periodic boundary conditions on the fields that are not present in type II, such as the tachyon. In particular, if $k$ is the winding number on the original circle (of radius $R$), then $k/2$ is the winding number on the circle of radius $2R$ and therefore the momentum on the T-dual circle. If $k$ is even, the string is a bosonic type II string; if $k$ is odd, the string is a type 0 string.

With this information in hand, we now apply Buscher duality \cite{Buscher:1987sk,Buscher:1987qj,Rocek:1991ps} to the $\theta$ direction of the geometry (\ref{pret}). We obtain the following NS-NS background:
\begin{equation}\label{postt}
ds^2 = dx^2+dr^2 + \left(\frac{N\alpha'}rd\tht\right)^2, \qquad
e^{2\Phi} = \frac{N^2\alpha'}{2r^2}e^{2\Phi_0}, \qquad
\tht\sim\tht + \pi.
\end{equation}
(The dilaton shift has an additional factor of $2$ compared to the usual Buscher rules due to the fact that the $\tht$ circle is half the size it usually is.) At finite $N$ there is a caveat: Buscher duality assumes that the resulting circle has a $U(1)$ isometry, which is the same as assuming that winding number $k$ is a conserved $U(1)$ charge on the original circle. In this case, the winding number symmetry is broken from $U(1)$ down to the ${\bf Z}_N$ quantum symmetry (whose charge is the twist number), due to the existence of the orbifold fixed point where $k$ can change by a multiple of $N$. Therefore the translational isometry in the $\tht$ direction of the geometry (\ref{postt}) is actually broken down to ${\bf Z}_N$; we can imagine $N$ copies of some unknown structure or defect evenly spaced around the $\tht$ circle at $r=0$ \cite{Adams:2001sv}.\footnote{An interesting fact about these defects is that D-branes can end on them. We deduce this from the fact that on \cn there are $N$ species of space-filling brane which are related to each other by the action of the quantum symmetry group. Upon T-duality these go to branes extended in $r$ and localized at $N$ possible values of $\tht$ that are related to each other by ${\bf Z}_N$ rotations. Furthermore, \cn has $N$ different species of fractional branes localized at the fixed point, also related to each other by the quantum ${\bf Z}_N$. These should be T-dual to branes extended from one structure to the next along the $\tht$ circle. This picture is consistent with the fact that if you have one of each kind of fractional brane then they can form a bound state which is a normal brane and move off the fixed point. It is also consistent with what happens when a space-filling brane and anti-brane annihilate, leaving behind a vortex in the tachyon field which is a fractional brane.} For the time being we will simply ignore them, assuming that at radii large compared to the string length their effect is negligible.

The Ricci scalar and dilaton gradient of the supergravity solution (\ref{postt}) blow up near the origin:
\begin{equation}
\mathcal{R} = -4\partial_M\Phi\partial^M\Phi = -\frac4{r^2}.
\end{equation}
The region in which we can use a low-energy ten-dimensional effective field theory on this background is therefore
\begin{equation}\label{validity}
\sqrt{\alpha'}\ll r < N\sqrt{\alpha'};
\end{equation}
below $\sqrt{\alpha'}$ the curvatures are large in string units, whereas above $N\sqrt{\alpha'}$ the winding modes become light, and a better description of the low energy dynamics is in terms of local fields on the original cone. At large $N$ the limits (\ref{validity}) still leave a large region in which to study the dynamics, including, as we shall see in the next subsection, that of the tachyons of interest.

\subsection{Tachyon spectrum from supergravity}

The twist number $k$ of a string (where again we take $0<k<N/2$) is equal to the number of times it winds around the $\theta$ circle, mod $N$. The reason for the ``mod $N$" is that, by passing through the fixed point, the string can gain or lose $N$ units of winding. However, we are interested in the tachyons with $k/N\ll1$, and these strings have a much larger amplitude for having winding number $k$ than any other allowed winding number, such as $k+N$ or $k-N$. This is for two reasons: first, this configuration has lower energy; second, as we will see below the wave function for this winding number is small (of order $k/N$) at the fixed point, suppressing the mixing between it and the other allowed winding numbers. We will therefore pretend that the tachyon is in a winding eigenstate with eigenvalue $k$; the error in this approximation is suppressed by powers of $k/N$.

As we saw in subsection \ref{thermalt} above, if $k$ is the winding number on the $\theta$ circle, then $k/2$ is the momentum on the $\tht$ circle. (Since $\tht$ has periodicity $\pi$, the wave function goes like $e^{ik\tht}$.) Furthermore, strings with odd $k$ are modes of the type 0 fields, while strings with even $k$ are modes of the (bosonic) type II fields. Specifically, for odd $k$ $T_k$ is the lowest energy mode of the type 0 tachyon field on the background (\ref{postt}) with $\tht$ dependence $e^{ik\tht}$; for even $k$ it is the lowest energy mode of the massless NS-NS fields with that $\tht$ dependence. We will now check that identification by explicitly finding those modes and reproducing the spectrum (\ref{spectrum}). Importantly, we will find that they are heavily concentrated inside the region (\ref{validity}) of validity of the T-dual field theory, so that the whole picture is self-consistent.

Call the type 0 tachyon field $t$. The effective 10-dimensional action for it, the metric, and the dilaton is (up to quadratic order in $t$)
\begin{equation}\label{action}
S =
\frac1{2\kappa^2}\int d^8\!x\,dr\,d\tht\,\sqrt{-G}e^{-2(\Phi-\Phi_0)} \left(
\mathcal{R} + 4\partial_M\Phi\partial^M\Phi
 - \partial_Mt\partial^Mt-m_t^2t^2
\right) +S_{\rm GH},
\end{equation}
where $m_t^2=-2/\alpha'$ and $S_{\rm GH}$ is the gravitational boundary term, which we will write when we need it later.

Let's start with the odd $k$ case, since it involves only the scalar field $t$. The linearized equation of motion for $t$ is
\begin{equation}
-\frac1{\sqrt{-G}e^{-2\Phi}}\partial_M\left(\sqrt{-G}e^{-2\Phi}\partial^Mt\right) + m_t^2t = 0.
\end{equation}
If we fix the $\tht$ dependence to be $e^{ik\tht}$ and look for eigenmodes of the two-dimensional Laplacian on the background (\ref{postt}), then we find the Schr\"odinger equation for a two-dimensional harmonic oscillator of frequency $k/N\alpha'$ and minimum $\frac12m_t^2$:
\begin{equation}\label{oddklap}
-\frac1r\partial_rr\partial_rt + \left(m_t^2+\left(\frac{kr}{N\alpha'}\right)^2\right)t = m_k^2t.
\end{equation}
This harmonic oscillator potential, which confines the string to the origin, is most easily understood on the original cone, where it is simply due to the tension of the string wound $k$ times around the $\theta$ circle. (Indeed, the two-dimensional harmonic oscillator of (\ref{oddklap}) is restricted to the $s$-wave because we are considering only strings with zero momentum on the $\theta$ circle.) The mode corresponding to $T_k$, the lowest tachyon in this winding sector, is the ground state of the harmonic oscillator, and the zero-point energy accounts  for the difference in mass between the 8-dimensional and 10-dimensional tachyons. The mode is
\begin{equation}\label{tfluc}
t(x,r,\tht) = \sqrt{2k}\exp\left(ik\tht-\frac{kr^2}{2N\alpha'}\right)T_k(x)
\end{equation}
(the normalization is chosen for convenience). Note in particular the width of the wave function, $\sqrt{N\alpha'/k}$, sitting comfortably between the two limits of (\ref{validity}). The probability of finding the string at $r<\sqrt{\alpha'}$ is roughly $k/N$ while the probability of finding it at $r>N\sqrt{\alpha'}$ is exponentially small (with respect to the measure $\sqrt{-G}e^{-2\Phi}$), bearing out the assertions of the previous subsections.

In the even $k$ case we are also looking for a mode with $\tht$ dependence $e^{ik\tht}$. Furthermore, the tension of the string wound
around the $\theta$ circle leads to the same harmonic oscillator potential as when $k$ is odd, so the $r$ dependence will be the same. However,
this tachyon is a fluctuation of the massless NS-NS modes, and the polarization can be found by applying Buscher duality to the vertex
operator. It turns out that it is a fluctuation of the metric only:
\begin{equation}\label{metricfluc}
\delta ds^2 =
\sqrt{2k} \exp\left(ik\tht-\frac{kr^2}{2N\alpha'}\right)\left(dr - i\frac{N\alpha'}rd\tht\right)^2T_k(x).
\end{equation}

The 8-dimensional effective action for the tachyons is found by expanding the gravity action (\ref{action}) in the fluctuations (\ref{tfluc}) and (\ref{metricfluc}) about the background (\ref{postt}). To quadratic order, this simply yields the free action,
\begin{equation}\label{taction}
S_2 =
\frac{2\pi\alpha'}{\kappa^2}\int d^8\!x
\sum_k\left(-\partial_\mu T_k\partial^\mu \bar T_k - m_k^2|T_k|^2\right).
\end{equation}
Because the fluctuation (\ref{metricfluc}) does not go to zero at $r=0$, to get the mass term right for the even $k$ tachyons it is essential to include the gravitational boundary term,
\begin{equation}\label{gh}
S_{\rm GH} =
\frac1{\kappa^2}\left.\int d^8\!x\,d\tht\,\sqrt{-\gamma}e^{-2(\Phi-\Phi_0)}K\right|_{r=0}^{r=\infty},
\end{equation}
where $\gamma$ and $K$ are the induced metric and extrinsic curvature of the boundary respectively. Expanding to higher order in the fluctuations yields interaction terms, as we will explore in the next section.

\section{Tachyon interactions}

\subsection{Generalities}

As explained in the previous section, the lightest (and heaviest) tachyons on \cn at large $N$ are well described as fluctuations of the metric (and bulk tachyon) in the T-dual background. Thus the gravity action (\ref{action}) provides a natural off-shell extension of string perturbation theory, as well as facilitating the computation of on-shell scattering amplitudes. In this section we will illustrate this procedure by calculating several tachyon couplings, starting in subsection 3.2 with the cubic and quartic couplings. In subsection 3.3 we will compute the cubic couplings in conformal field theory, and compare them to the gravity results. Finally, in the last subsection we will use the gravity action to reproduce the CFT results of \cite{Dabholkar:2004} concerning couplings between the tachyons and massless fields on ${\bf C}/{\bf Z}_N$.

Before launching into the calculations let us make two general comments regarding the interactions. First, the power of $N$ appearing in each coupling constant is essentially fixed by dimensional analysis, as follows. The background (\ref{postt}) is characterized by the length scale $\sqrt{N\alpha'}$; $N$ and $\alpha'$ do not appear separately (aside from an extra factor of $N$ in the dilaton, which gets absorbed in the overall prefactor in the action). Furthermore the 10-dimensional action (\ref{action}) has no dimensionful or dimensionless parameters (aside from $\kappa$, which again only appears in the prefactor, and $m_t^2$, which will affect only certain couplings of the odd-$k$ tachyons). Therefore the power of $N$ appearing in a coupling constant must be the same as the power of $\alpha'$, which in turn is determined by dimensional analysis. Specifically, in view of the fact that the integrand in the 8-dimensional action (see (\ref{taction})) has dimensions of mass-squared, and that the tachyon fields $T_k$ are dimensionless, each term in the potential will go like $(N\alpha')^{-1}$ and each derivative coupling will go like $(N\alpha')^0$.

The second comment is that results derived from gravity, such as in this section, are valid only at leading order in $1/N$, and will receive subleading corrections from a variety of sources. For example, the low-energy action (\ref{action}) is not valid in the region $r<\alpha'$, where the tachyons of interest have a probability about $1/N$ of being found. And since we are working at fixed $N\alpha'$, $\alpha'$ corrections to the Buscher duality rules and to the low-energy action (\ref{action}) will yield further $1/N$ corrections. Obviously we will not attempt to compute any of these subleading terms.

\subsection{Cubic and quartic interactions from supergravity}

For cubic tachyon couplings $T_{k_1}T_{k_2}\bar T_{k_1+k_2}$, we can distinguish three cases: (1) $k_1$ is odd and $k_2$ is even, or vice versa; (2) $k_1$ and $k_2$ are both odd; (3) $k_1$ and $k_2$ are both even. In the first two cases the interaction arises from the kinetic term for $t$ in the action (\ref{action}). When expanded to first order in the metric fluctuation this gives $h^{MN}\partial_Mt\partial_Nt$, where $h_{MN}$ is the metric fluctuation (\ref{metricfluc}) (note that $h_{MN}$ is traceless, so there is no term from expanding the measure). However, the polarization of the metric fluctation is such that $h^{MN}\partial_Mt$ vanishes when $h_{MN}$ and $t$ both have positive $k$. Hence in case (1) the coupling vanishes. There seems to be some kind of helicity selection rule at work, presumably the spacetime manifestation of the $H$-charge conservation that (as noted in the next subsection) makes this coupling vanish when computed by the world-sheet method. In case (2) the coupling does not vanish, and we have the following action:
\begin{equation}\label{cubic}
S_3 = \frac{2\pi\alpha'}{\kappa^2}\int d^8\!x
\sum_{k_1,k_2\,{\rm odd}}
\frac1{N\alpha'}\left(\frac{2k_1k_2}{k_1+k_2}\right)^{3/2}T_{k_1}T_{k_2}\bar T_{k_1+k_2}
+ {\rm c.c.}
\end{equation}
It turns out that in case (3) the coupling vanishes, as one can show by expanding the metric-dilaton action to third order in the metric fluctuations (a task for which one is greatly aided by {\it Mathematica}). For this it is important to include the boundary term (\ref{gh}). Again, the helicity selection rule seems to be at work.

To compute the quartic coupling for the odd-$k$ tachyons we would need to know the quartic term in the 10-dimensional action for the type 0 tachyon. However, for the even-$k$ tachyons, the metric-dilaton part of (\ref{action}) allows one to calculate arbitrarily high order interactions in principle. Their quartic coupling, unlike their cubic coupling, does not vanish; in fact it includes a derivative interaction, making the field space curved:
\begin{equation}\label{quartic}
S_4 = \frac{2\pi\alpha'}{\kappa^2}\int d^8\!x
\sum_k\left(-2k|T_k|^2\partial_\mu T_k\partial^\mu\bar T_k +\frac{6k^2}{N\alpha'}|T_k|^4\right).
\end{equation}
We have only calculated the self-interactions, but presumably there are also quartic interactions among tachyons with different values of $k$.

\subsection{Cubic interactions from conformal field theory}

We will now compute the cubic tachyon couplings in string perturbation theory and show that they agree, in the large $N$ limit, with the supergravity result (\ref{cubic}).\footnote{The results of this subsection were also obtained independently by Y. Okawa \cite{Yuji}, whom we would like to thank for useful discussions on this subject.} The amplitude is given by a correlator
\begin{equation}
\mathcal{M} = \langle V_{k_1} (\infty ) V_{k_2}(1) \bar V_{k_1 + k_2}(0)\rangle
\end{equation}
where the $V_{k_i}$ denote the vertex operators corresponding to the tachyonic fields $T_{k_i}$. We need to take two vertex operators in the $-1$ picture and one vertex operator in the $0$ picture; by a judicious choice of pictures we will find that the amplitudes found to vanish in the previous subsection vanish here because of $H$-charge conservation. In the $-1$ picture, the vertex operators for scalar tachyons ($k$ odd) and tensor tachyons ($k$ even) are given by (see \cite{Dabholkar:2004} for more details)
\bea
V_{k}^{(-1)} &=& \frac{\k \sqrt{N} }{ 2 \p }\s_k e^{ik(H -\tilde H)/N} e^{ip_\m X^\m} c \tilde c e^{-(\f + \tilde \f)} \qquad \qquad
\hbox{$k$ odd; scalar tachyons} \\
V_{k}^{(-1)} &=& \frac{\k \sqrt{N} }{ 2 \p} \s_k e^{i(k-N)(H -\tilde H)/N} e^{ip_\m X^\m} c \tilde c e^{-(\f + \tilde \f)} \qquad
\hbox{$k$ even; tensor tachyons.}
\eea
The normalization of these vertex operators was determined in \cite{Dabholkar:2004} from unitarity.\footnote{Note that the constant $\k$ in section 3 of that paper is related to the one used here by $\k_{\rm there} = \k_{\rm here} \sqrt{N}$.} The vertex operators in the zero picture are obtained by acting with modes of the matter supercurrent: $V_{k}^{(0)} = G_{-1/2} \tilde
G_{-1/2} V_{k}^{(-1)}$.

As in the previous subsection, we distinguish three cases: (1) If $k_1$ is odd and $k_2$ is even, taking $V_{k_1}$ in the zero picture one finds that the amplitude vanishes due to $H$-charge conservation, in agreement with the gravity result. (2) If $k_1$ and $k_2$ are both odd, the amplitude is allowed by $H$-charge conservation. Taking $\bar V_{k_1 + k_2}$ to be in the 0 picture the amplitude  is given by a correlator in the orbifold CFT involving an excited twist field:
\begin{equation}\label{CFTc's}
\mathcal{M} = - \frac{4 \kappa \sqrt{N}}{\a'} \left\langle
\sigma_{k_1}'(\infty ) \sigma_{k_2}(1) \alpha_{-1 + (k_1 +
k_2)/N} {\bar{\tilde \alpha}}_{-1 + (k_1 + k_2)/N} \cdot
\sigma_{N-k_1 -k_2}(0)\right\rangle.
\end{equation}
The correlator can be computed from the 4-twist correlator $\langle\sigma_{k_1}\sigma_{-k_1}\sigma_{k_2} \sigma_{-k_2}\rangle$ by factorizing on the intermediate state with the proper conformal weight \cite{Dixon:1987qv}. For example, for the case $k_1 = k_2 = k$ one finds
\begin{align}
&\left\langle
\sigma_{k}'(\infty)
\sigma_{k}(1) \alpha_{-(1 - 2k/N)} {\bar{\tilde \alpha}}_{-(1 - 2k/N)} \cdot \sigma_{-2k}(0)
\right\rangle \nonumber \\
& \qquad\qquad\qquad\qquad\qquad\qquad\qquad =
\frac{(1- 2 k/N)}{\sqrt{2 \pi^2 \alpha '}} \sqrt{\tan \frac{\pi k}N} \left(\frac{\Gamma^2(1 -  k/N)}{\Gamma(1 - 2k/N)}\right)^2
\frac{\Gamma(2 k/N -1)}{\Gamma^2(k/N)} \nonumber \\
& \qquad\qquad\qquad\qquad\qquad\qquad\qquad =
-\frac1{2 \sqrt{2\pi\alpha'}} \left(\frac kN\right)^{3/2}\left(1+\mathcal{O}\left(\frac1N\right)\right).
\end{align}
Substituting into (\ref{CFTc's}) gives
\begin{equation}
\mathcal{M} = \frac{2\kappa k^{3/2}}{\sqrt{2\pi}N\alpha^{\prime3/2}},
\end{equation}
which agrees with the amplitude calculated from the action (\ref{taction}, \ref{cubic}). (3) If $k_1$ and $k_2$ are both even one finds that, again taking the vertex operator $\bar V_{k_1 + k_2}$ in the 0 picture, the amplitude vanishes due to $H$-charge conservation, in agreement with the gravity result.

\subsection{Interactions with massless fields}

So far we have concentrated on tachyon-tachyon interactions. However, it is also straightforward to use the T-dual picture to calculate couplings between tachyons and other fields. In \cite{Dabholkar:2004} the amplitude for scattering of two tachyons and a massless field on \cn was computed using string perturbation theory. Here we will reproduce this result using the action (\ref{action}). In addition, we will find extra couplings that vanish when all fields are on shell.

A mode expansion on the cone (\ref{pret}) for the massless NS-NS supergravity fields (metric, $B$-field, dilaton) may be obtained by taking a plane wave on the covering space and averaging over all rotations about the origin, to yield modes that are independent of $\theta$. There are other modes that have momentum in $\theta$, but at large $N$ they are very massive from a $7+1$ dimensional point of view and upon T-duality go into winding modes on $\tht$, hence we neglect them.\footnote{Strictly speaking, rather than averaging over all rotation angles, one should average only over multiples of $2\pi/N$ (as in (3.2) of \cite{Dabholkar:2004}). Fourier expanding the result in $\theta$, the terms with non-zero $\theta$-momentum have coefficients of order $1/N!$, so we neglect them.} Following \cite{Dabholkar:2004}, on the covering space we use coordinates $x^\mu,X=r e^{i\theta}/\sqrt{2},\bar X=r e^{-i\theta}/\sqrt{2}$. Without loss of generality we take the $X$-momentum real, $p_X=p_{\bar X}$. The graviton is given by the symmetric part of the polarization tensor, the $B$-field by its antisymmetric part, and the dilaton by its trace (the tildes indicate that these are modes on the cone (\ref{pret}) rather than on the T-dual geometry (\ref{postt})):
\begin{equation}\label{gauge1}
\tilde h_{MN} = e_{(MN)}e^{ip_Mx^M}, \qquad
\tilde b_{MN} = e_{[MN]}e^{ip_Mx^M}, \qquad
\tilde \phi = \frac14{e_M}^Me^{ip_Mx^M}.
\end{equation}
The polarization tensor must be transverse to the momentum:
\begin{equation}\label{gauge2}
p^Me_{MN} = p^Ne_{MN} = 0.
\end{equation}

After averaging over rotation angles and applying Buscher duality, the modes on the T-dual geometry (\ref{postt}) are as follows:
\begin{align}
h_{\mu\nu} &= e_{(\mu\nu)} j_0 \\
h_{\mu r} &= \frac{i}{\sqrt2} (e_{(\mu X)}+e_{(\mu\bar X)}) j_1 \\
h_{\mu\tht} &= \frac{N\alpha'}{\sqrt2r} (e_{[\mu X]}-e_{[\mu\bar X]})j_1 \\
h_{rr} &= e_{(X\bar X)} j_0 - \frac12 (e_{XX}+e_{\bar X\bar X}) j_2 \\
h_{r\tht} &= -\frac{iN\alpha'}re_{[X\bar X]}j_2 \\
h_{\tht\tht} &=
-\frac{N^2\alpha'^2}{r^2} \left(e_{(X\bar X)}j_0 + \frac12(e_{XX}+e_{\bar X\bar X}) j_2 \right) \\
\phi &= \frac14 {e_\mu}^\mu j_0 - \frac14 (e_{XX}+e_{\bar X\bar X}) j_2
\end{align}
(The modes of the T-dual $B$-field are unnecessary for our purposes, since they don't couple at first order to the tachyons.) The wave functions depend on $r$ through Bessel functions:
\begin{equation}
j_i = J_i\left(\sqrt2p_Xr\right)e^{ip_\mu x^\mu}.
\end{equation}

Plugging these modes into the action (\ref{action}) together with the tachyon modes (\ref{tfluc}) and (\ref{metricfluc}) and expanding to first order in the massless modes and second order in the tachyons, one obtains the following scattering amplitude:
\begin{equation}\label{ttmassless}
\mathcal{M} =
 2\kappa\sqrt{N}e^{-N\alpha'p_{3X}^2/2k}
\left(p_2^\mu p_2^\nu e_{\mu\nu}-p_2^\mu p_{3X}(e_{\mu X}+e_{\bar X\mu})+p_{3X}^2e_{\bar XX}\right).
\end{equation}
Here $p_3$ is the momentum of the massless field and $p_2$ is the momentum of $\bar T_k$ (for odd $k$) or $T_k$ (for even $k$). The result (\ref{ttmassless}) agrees with the large-$N$ limit of the same amplitude as calculated in CFT, eq.\ (4.12) of \cite{Dabholkar:2004}.

So far we have only considered on-shell amplitudes. The supergravity action  also gives a consistent off-shell extension and one expects it to be able to answer off-shell questions which are hard to address using on-shell string amplitudes. For example, in the case of the 3-point couplings, one finds that there are additional contributions that vanish when the particles are on-shell. These contributions are different for the scalar and tensor modes. We will presently examine those terms in more detail.

When the fields are off their mass shells, it is no longer possible reach a gauge where the transversality condition (\ref{gauge2}) and and the condition on the trace $\tilde\phi ={\tilde h_M}^M/4$ in (\ref{gauge1}) are satisfied simultaneously. We will therefore relax the condition $\tilde\phi = {\tilde h_M}^M/4$, allowing the dilaton to be an independent degree of freedom
from the metric. The transversality condition (\ref{gauge2}) on the graviton correspondingly gets replaced by
\begin{equation}
\nabla^M \tilde h_{MN} - \half \nabla_N {\tilde h_M}^M + 2 \nabla_N \tilde\phi = 0,
\end{equation}
while that on the $B$-field remains $\nabla^M\tilde b_{MN}=0$ as before.
For the coupling of the scalar tachyons to gravitons, one finds that  there is an extra off-shell coupling in addition to (\ref{ttmassless}):
\begin{equation}\label{scalarttgrav}
\d \mathcal{M}^{\rm grav} =  \k \sqrt{N}e^{-N\alpha'p_{3X}^2/2k} (p_2^2 + m_k^2) {e_M}^M \end{equation}
as well as a coupling to the dilaton:
\begin{equation}\label{scalarttdil}
\d \mathcal{M}^{\rm dil} =  - 4 \k \sqrt{N}e^{-N\alpha'p_{3X}^2/2k} (p_2^2 + m_k^2).
\end{equation}
For the tensor tachyons there are further contributions in addition to (\ref{scalarttgrav}, \ref{scalarttdil}). One finds  an extra off-shell coupling to the graviton:
\begin{multline}\label{tensorttgrav}
\d \mathcal{M}^{\rm grav} =
 - 2 \k \sqrt{N} e^{-N\alpha'p_{3X}^2/2k}  p_3^2 {e_M}^M \\
 + 2 \k \sqrt{N}\left(e^{-N\alpha'p_{3X}^2/2k}\left(\frac{2k}{N\alpha'p_{3X}^2}+1\right)-\frac{2k}{N\alpha'p_{3X}^2}\right)
\left(p_1^2+p_2^2+2m_k^2+p_3^2\right)
\left(e_{XX}+e_{\bar X\bar X}\right),
\end{multline}
and a coupling to the dilaton
\be\label{tensorttdil}
\d \mathcal{M}^{\rm dil} = 8 \k \sqrt{N}e^{-N\alpha'p_{3X}^2/2k}p_3^2.
\ee

Following standard field theory lore, one would expect that the off-shell terms (\ref{scalarttgrav}--\ref{tensorttdil}) can be removed by making suitable nonlinear field redefinitions. This was implicitly assumed in \cite{Dabholkar:2004}, where (\ref{ttmassless}) was used as the off-shell 3-point vertex. There is however a subtlety to this procedure because the tachyonic perturbations (\ref{tfluc}, \ref{metricfluc}) do not vanish on the boundary $r=0$. For example, to remove the first term in (\ref{tensorttgrav}) and (\ref{tensorttdil}), one would have to shift the mode $\tilde h_M^M - 4 \tilde \phi$ by an amount proportional to $ T_k \bar T_k e^{- kr^2/N \a'}$. With the present choice of boundary term, this field redefinition affects the mass term for the tachyon and the other cubic couplings (\ref{cubic}). One could avoid  this difficulty by using a non-standard boundary term containing terms proportional to $\tilde h_M^M - 4 \tilde \phi$ in addition to (\ref{gh}), but this seems rather unnatural.

\acknowledgments

We would like to thank A. Adams, S. Minwalla, Y. Okawa, E. Silverstein, and especially A. Dabholkar and B. Zwiebach for helpful discussions. This work was initiated while we were Visiting Fellows in the Department of Theoretical Physics of the Tata Institute of Fundamental Research. We would also like to thank Harvard University, SLAC, and Stanford University for hospitality. M.H. is supported by a Pappalardo Fellowship, and also by funds provided by the U.S. Department of Energy under cooperative research agreement DF-FC02-94ER40818.

\bibliography{ref}
\bibliographystyle{JHEP}

\end{document}